\documentclass[10pt,letterpaper,twocolumn]{article}
\usepackage[english]{babel}
\usepackage{graphicx}

\newcommand{\alt}{\mathbin{\lower 3pt\hbox
   {$\rlap{\raise 5pt\hbox{$\char'074$}}\mathchar"7218$}}}
\newcommand{\agt}{\mathbin{\lower 3pt\hbox
   {$\rlap{\raise 5pt\hbox{$\char'076$}}\mathchar"7218$}}}

\begin{document}

\setcounter{footnote}{0}
\setcounter{equation}{0}
\setcounter{figure}{0}
\setcounter{table}{0}

\title{\large\bf
Reply to comment by P.\,Markos [arXiv:1205.0689] }

\author{\small I. M. Suslov \\
\small P.L.Kapitza Institute for Physical Problems,
Moscow, Russia \\ {}\\
\footnotesize We present another interpretation of the data by
P.\,Markos and give a lot of  new    \\
\footnotesize \,illustrations
for our conception. All existing
numerical data look perfectly
compa-
\\  \footnotesize \quad tible with
predictions of the self-consistent theory of  localization.
\qquad\qquad\qquad\qquad\qquad
}

\date{}

\maketitle

\begin{figure*}
\centerline{\includegraphics[width=3.5 in]{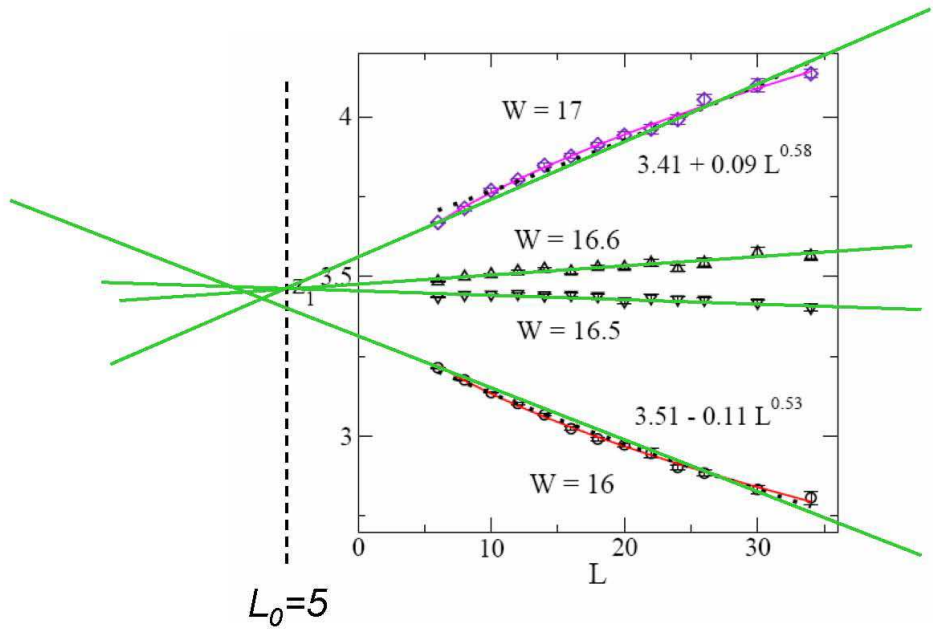}}
\caption{\small  3D data for $z_1=2L/\xi_{1D}$  \cite{4} and
their fitting by dependence $C(L+L_0)$.} \label{fig1}
\end{figure*}

Our paper \cite{1} contains the detailed predictions of the
self-consistent theory of localization  for the quantities
which are immediately measured in numerical experiments;
it allows to make a comparison  on the level of the raw data,
avoiding the ambiguous treatment procedure.  Such approach
is motivated by the different status of numerical results.
The raw data are obtained independently by different groups and
there is a certain consensus in this respect; it is not
reasonable to question these data. However, it is
possible to doubt numerical algorithms themselves, which are not
based on a firm theoretical ground.
Such approach is in the own interests of  numerical researches
since their present-day  results contradict
both experiment and the general theoretical principles.
Self-consistent theory by Vollhardt and W${\rm {\ddot
o}}$lfle allows to justify (for the first time)  one of the
popular variants of finite-size scaling based on consideration
of auxiliary quasi-1D systems \cite{2,3} with a finite
transverse size $L$. This theory  predicts also the essential
scaling corrections, so the scaling parameter has a behavior $C
(L+L_0)$ with $L_0>0$ in the vicinity of transition, which can be
practically interpreted as $C L^{1/\nu}$ with $\nu>1$.
Consideration of existing numerical data  shows that there are no
serious contradictions of the self-consistent theory with the raw
numerical data.

Of course, it does not prove a validity of the self-consistent
theory:  deviations can be  small but significant, and a
serious
analysis is necessary. The analysis of this
kind is expected from the specialists in numerical research,
such as P.\,Markos.
In fact, in the comment \cite{4} he makes no efforts
to follow  our suggestions but restricts himself
by the  "standard scaling
formulas". First of all, there are no
"standard scaling formulas", since corrections to
scaling certainly exist  and  no reliable procedure
to deal with them is available. Further, the conventional
scaling is certainly invalid for dimensions $d>4$: it is a
theorem \cite{1}.  Finally, we did not deny in \cite{1}
the possibility to fit the data by a simple power
law dependence  but  stressed
ambiguity of such procedure.  From this point of view, FIGS.\,2--5
in \cite{4} have no relation to the criticism of the paper
\cite{1}.

\vspace{5mm}

\begin{figure*}
\centerline{\includegraphics[width=5.1 in]{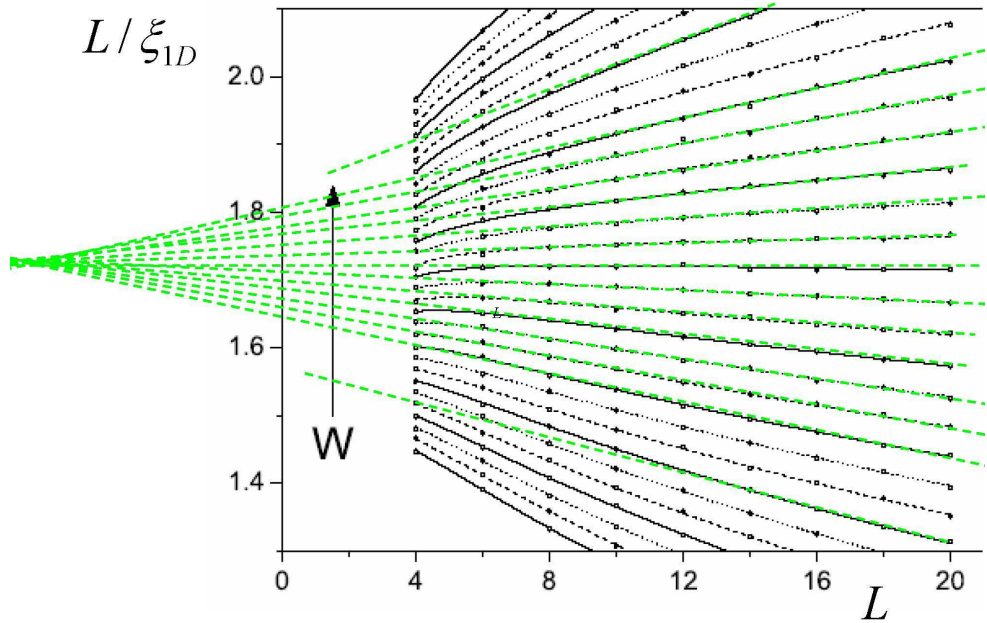}}
\caption{\small The same, as in Fig.\,1, with
smaller $L$ and higher accuracy (from the paper
by Kramer et al \cite{208}). For small deviations from the
critical point the data are fitted well by dependence
$C(L+L_0)$. When deviations become large, dependencies
acquire essential curvature,
while   effective $L_0$ changes significantly.  Compare with
Fig.\,1.} \label{fig2} \end{figure*}

{\bf 3D system}. In this case, P.\,Markos provides not a very
essential progress: he extends his results till $L=34$, while
data up to $L=50$ were discussed in \cite{1}.
Our interpretation of 3D data is presented in Fig.\,1. The
following points should be noted:

(a) The most interesting question is: does  $L_0$ reveal
an essential drift when the range of $L$ is extended? If we try
to retain the estimate $L_0=5$ obtained in \cite{1} for $L \le
24$, then the data for $W=16.5$ and $16.6$ are
fitted well with such restriction.

(b) The data for $W=16$ and $W=17$ show certain deviations from
the linear behavior  but they
 are not very impressive, since the
scattering of points is rather large.

(c) In fact, the data for $W=16$ and $W=17$  contain the
effect of the $W$ nonlinearity. If we suggest $\nu=1$, then
$\xi\approx 30$ for $|W-W_c| =0.5$ and  nonlinear
effects are  essential for $L\sim 30$. Fig.\,1  confirms this
conclusion, since the data  for $W=16$ and $W=17$ are not
symmetric relative to the curve
$W=16.5$.\,\footnote{\,In fact,
Fig.\,1 roughly confirms that $\xi\approx 30$, since deviations of
$z_1$ from its critical value is of the order of unity (if
$\nu=1.5$ then $\xi$ should be something like 150).}
Deviations from the linear behavior are on the same
level as violation of symmetry.
It looks rather probable  that for the more narrow interval
(like $W=16.25 \div 16.75$)  fitting by  the linear
dependence will be satisfactory.\,\footnote{\,It is clear from
FIG.\,2 in \cite{4} that the author has the intermediate data for
Fig.\,1. Why he does not show them? } This argumentation is
supported by other numerical data (Fig.\,2).

\vspace{3mm}
\begin{figure*}
\centerline{\includegraphics[width=4.0 in]{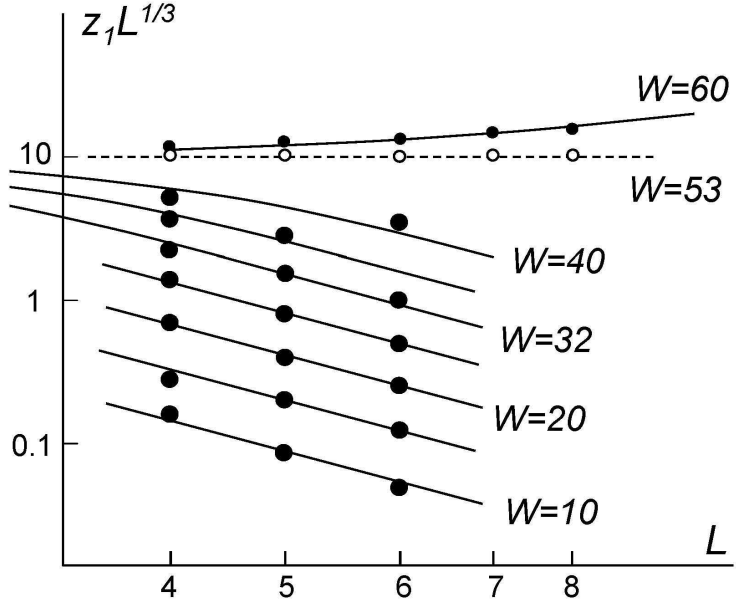}}
\caption{\small 5D data for $z_1=2L/\xi_{1D}$ extracted from
FIGS.\,4,\,5 in \cite{4}
(points) and their comparison with the scaling relation (2)
(solid lines).} \label{fig3} \end{figure*}

P.\,Markos has an illusion that the more complicated procedure
allows to obtain the higher accuracy. In particular,
 in treatment of
the $W$ dependence he relies on the
quadratic expansion in $W-W_c$.  In fact, one cannot exclude
possibility that the coefficient of the quadratic term is small
and the higher order corrections are essential. If different
nonlinear functions are allowed, the  uncertainty will be the
same  as for a simple linear fit in the more narrow interval.
In the latter case, it is impossible to obtain a nonlinear
behavior for the derivative $s(L)=\left[ z_1(L) \right]'_\tau $
from the apparently linear dependencies  $z_1(L)$ (Fig.\,2).
With nonlinear treatment, P.\,Markos was able
to do it (see FIG.\,3 in \cite{4}).

Comparison in FIG.\,3 of \cite{4} is not honest, since the dashed
line does not correspond to predictions of  \cite{1}. The
predicted dependence is  $C(L+L_0)$ and not $C L$, so the
straight line with the unit slope
is irrelevant. In fact, our concept works excellently in the
range $L\le 20$ (Fig.\,2), where P.\,Markos shows disastrous
deviations.

\vspace{5mm}

{\bf 5D model}. In this section we read:
\vspace{3mm}

\parbox{7cm}{\small
"Our data in FIG.\, 4 do not indicate
any discontinuity in the $L$ dependence. Contrary, $z_1$ is
smooth analytical function of both parameters, $W$ and
$L$."}

\vspace{3mm}
\noindent
We do not predict any discontinuity, it is a
fantasy by P.\,Markos.
It is clear from Eq.45 in \cite{1}
$$
\tau \Lambda^{d-2} =
\frac{1}{L^{d-2}}\, \frac{1}{2 m L} - c m^2 \Lambda^{d-4}
$$
that $mL\equiv z_1$ is a regular function of $L$ and
$\tau=W-W_c$.  A singularity is developed only in the
thermodynamic limit $L\to\infty$, as in all scaling theories.
Modifications suggested  for $d>4$ correspond to the usual
scaling constructions, but  in  other variables
$$
y=\frac{\xi_{1D}}{L} \left(\frac{a}{L} \right)^{(d-4)/3}\,,\qquad
x=\frac{\xi}{L} \left(\frac{a}{L} \right)^{(d-4)/3}\,.
\eqno(1)
$$
The scaling relation is found in the analytical form
$$
\pm \frac{1}{x^2} = y -\frac{1}{y^2} \,,
\eqno(2)
$$
where the proper scales for $\xi_{1D}$ and $\xi$ are chosen.
Fig.\,3 shows the quantity  $z_1 L^{1/3}\equiv 1/y$  as a function
of $L$.  Its dependence  on $1/x\sim L^{4/3}$ has the same form
but the logarithmic scale should be changed by the factor $4/3$.
The solid lines correspond to the scaling relation (2).
According to Fig.\,3, the critical point is $W_c=53$ and not
57.5.

\vspace{5mm}

{\bf Conclusion.} In this section,  the author provides the
additional argumentation:

 \vspace{3mm}
\parbox{7cm}{\small
"We also note that the same value of the critical
exponent was obtained from numerical analysis of other
physical quantities: mean conductance, conductance
distribution, inverse participation ratio..."}
\vspace{3mm}

\noindent
In fact, two variants of scaling, (a) quasi-1D systems, and
(b) level statistics, were discussed in \cite{1}.  The third
variant, (c) mean conductance, is discussed in the recent paper
\cite{5}. The rest two variants, (d) conductance distribution
\cite{10}, and (e) inverse participation ratio \cite{11},
are illustrated in Figs.\,4 and 5. One can see, that our
conception is supported by high-precision data with $L\le 20$
and by moderate-precision data with $L\le 50$.

\begin{figure*}
\centerline{\includegraphics[width=5.9 in]{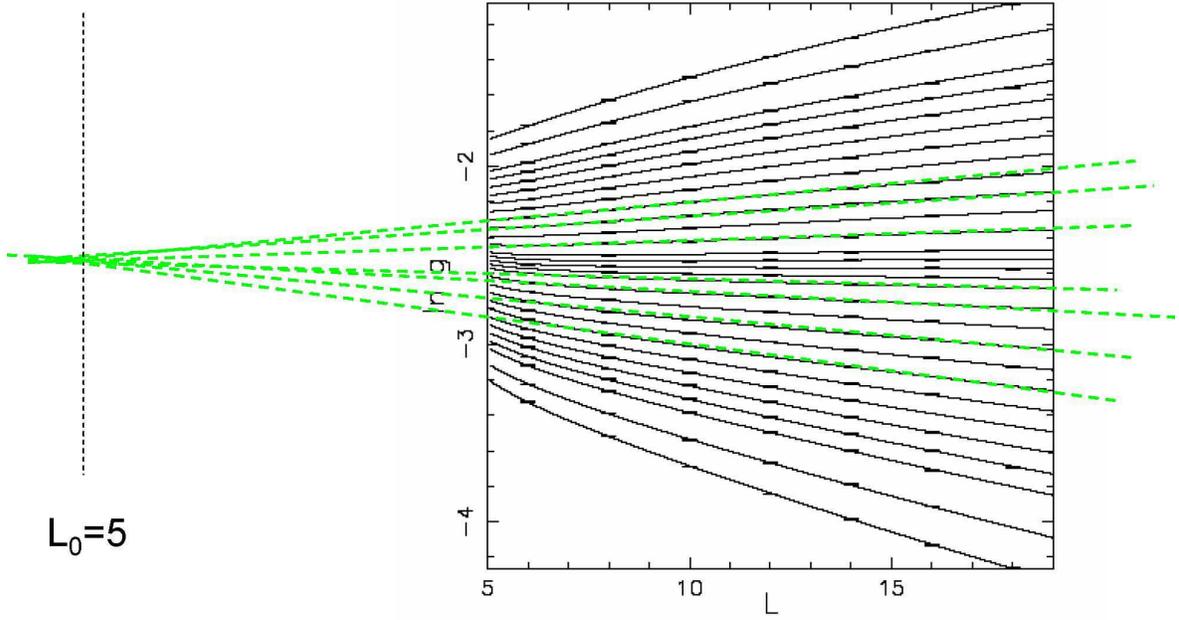}}
\caption{\small Data for the conductance distribution \cite{10}
and their fitting by dependence $C(L+L_0)$. The scaling parameter
is the 0.17 percentile of distribution.} \label{fig4}
\end{figure*}

\begin{figure*}
\centerline{\includegraphics[width=4.4 in]{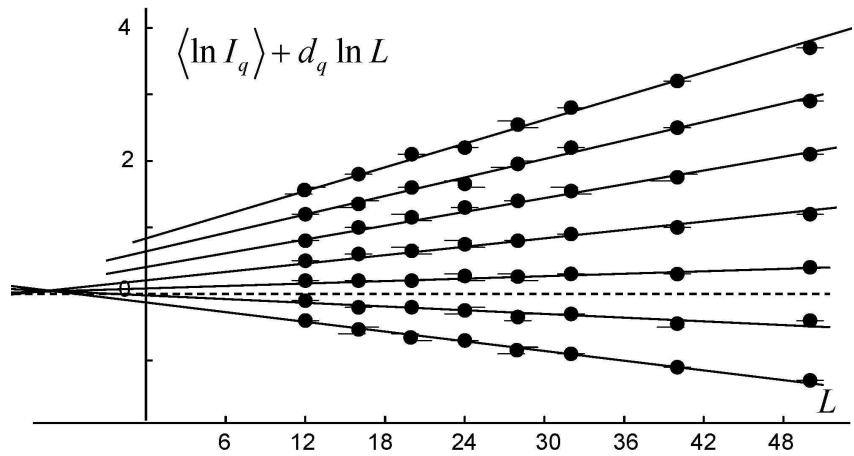}}
\caption{\small Data for the inverse participation ratio
$I_q$ with $q=5$ extracted from Fig.2 in \cite{11} and their
fitting by dependence $C(L+L_0)$.  The essential
change of $L_0$ is visible for large deviations from the
critical point. The contribution $-d_q\ln L$ ($d_q$ is a fractal
dimension) corresponds to the critical point.} \label{fig5}
\end{figure*}

\vspace{1mm}

The final arguments are also not serious:
\vspace{3mm}

\parbox{7cm}{\small
"This value of the
critical exponent was recently
verified experimentally ${}^{[11]}$ and calculated
analytically ${}^{[12]}$"}

\vspace{3mm}
The papers  ${}^{[11]}$ deal with a quasiperiodic kicked rotor,
whose equivalence with the 3D Anderson model is only a hypothesis
essentially based on the questinable numerical
data.\,\footnote{\,In fact, localization in quasiperiodical
systems has essential specificity in comparison with random
systems \cite{12}.} The real experiments on disordered systems
\cite{6,7,8} support the results of the self-consistent theory.

The "analytical" result ${}^{[12]}$ violates the Wegner scaling
relation $s=\nu(d-2)$, which is admitted by all serious
theoreticians.  Its violation means incorrectness of the
one-parameter scaling hypothesis \cite{9}, which is a basis
for practically all numerical studies.

\vspace{3mm}

In conclusion, P.\,Markos does not see the central idea of the
paper \cite{1} and continue to use the sophisticated treatment
procedure
instead of direct comparison on the level of raw data. If the
latter is made, all existing numerical data look perfectly
compatible with predictions of the self-consistent theory of
localization.

\end{document}